# Plutonium and Quantum Criticality

G. Chapline, M. Fluss, and S. McCall

Lawrence Livermore National Laboratory, Livermore, CA 94551

E-mail: chapline1@llnl.gov

**Abstract.** The unusual properties of the elemental plutonium have long been a puzzle. It has been suggested that these properties may be related to quantum criticality [1]. In this talk we will describe some experimental observations on rare earth and actinide materials which suggest that there are pairing correlations in all *f*-electron metals, and that the anomalous properties of the actinides in the vicinity of Np/Pu/Am, even at elevated temperatures, is associated with a critical point in the variation of the density of paired *f*-electrons with atomic number.

At its onset the Manhattan Project was bedeviled with a peculiar problem; the lattice structure and density of elemental Pu seemed to have two different values, depending on how it was prepared [2]. Perhaps even more remarkably, despite the importance of this problem and the passage of more than 60 years, understanding why Pu occurs in these two allotropic forms has proved elusive. Another mystery is why Pu is not magnetic. By the usual atomic rules it is expected that the core *f*-electrons of Pu ions should have a non-vanishing angular momentum and associated magnetic moment. However, there is considerable evidence that there are no local magnetic moments in solid Pu [3]. An intertwined puzzle is that despite the absence of magnetism, both Np and Pu have anomalously large magnetic susceptibilities [4].

Fortunately a solution to these puzzles may be on the horizon; namely elemental Pu lies close to a quantum critical point (QCP) of a particular kind. One hint that Pu lies close to a QCP of some kind is the case is that as a function of relatively modest changes in temperature and pressure there is a proliferation of complex phase transitions in Np, Pu, and Am that involve changes in the lattice structure and possibly also the electronic wave function of the 5f electrons [5]. Consideration of atomic volumes suggests that it is impurity stabilized δ Pu that may be nearest to the QCP point. Indeed, the negative thermal expansion of δ Pu for low Ga doping is by itself indicative of critical behavior. Another hint that there is a QCP in the vicinity of Pu or Am is that the ground state of the elemental actinides below Cm have non-magnetic ground states, while the transuranic elements beyond Am have anti-ferromagnetic ground states (Fig. 1). A quantum phase transition of the ground state from being anti-ferromagnetic and metallic to a normal Fermi liquid is quite typical low temperature behavior for many *f*-electron materials at high pressures or as their stoichiometry is changed [6]. The overall impression in the case of the elementary actinides under pressure or as the atomic number is increased is that the quantum ground state undergoes some kind of quantum phase transition in the vicinity of Pu. It is intriguing in this connection that the existence of a condensed matter QCP at zero temperature often results in unusual properties at ordinary temperatures [7]. In quantum statistical mechanics the temperature acts as an infrared cutoff for the quantum fluctuations rather than a tuning parameter. Therefore, if one is tuned to a QCP at low temperature increasing the temperature will have little effect if the quantum fluctuations are localized in space. It is probably not correct to postulate that the quantum fluctuations are localized to atomic sites, as is typically assumed in DMFT calculations [8]. However, the existence of a QCP with quasi-localized quantum fluctuations may well explain the anomalous properties of Np/Pu/Am at all temperatures up to their melting temperatures.

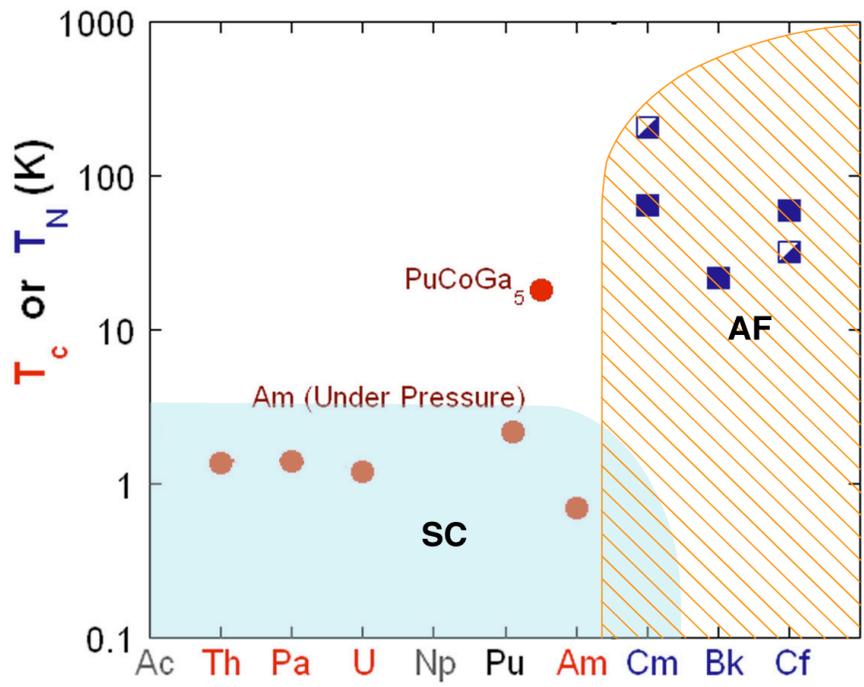

Fig 1. Magnetism vs. superconductivity in the actinides.

An obvious question that arises in this connection is what is the nature of the changes in the ground state of the elemental actinides that is responsible for the quantum critical behavior? The change from a non-magnetic ground state to an anti-ferromagnetic ground state with increasing atomic number as illustrated in Fig. 1 must surely be a clue. The occurrence of exceptional superconductivity in $PuCoGa_5$ and Am under pressure is probably also worth noting. In general the intensity of anti-ferromagnetic fluctuations in transition metal compounds is closely correlated with the transition temperature for exceptional superconductivity over an enormous range of materials ranging from heavy electron materials to the high $T_c$ cuprates [9]. Our guess is that the ground state of the elemental actinides may be very similar to the pseudo-gap phase of the high $T_c$ cuprates. In both cases there is no bulk superconductivity, although there may be a pair condensate. The idea of having a pair condensate without bulk superconductivity (a "gossamer superconductor") was originally put forward by Bob Laughlin in order to explain the transition between anti-ferromagnetism and superconductivity in the high $T_c$ cuprates [10]. In the gossamer picture the pairing correlations responsible for superconductivity can coexist with anti-ferromagnetism. In the case of either the hole or electron doped cuprates the gossamer condensate will persist down to zero doping, even for a doping where bulk superconductivity has disappeared. That such a picture might also apply to *f*-electron materials is encouraged by the discovery that as $CeRhIn_5$ changes from being anti-ferromagnetic to superconducting as the pressure is increased, there is a range of pressures where superconductivity and anti-ferromagnetism coexist [11].

Subsequent to Laughlin's suggestion it was suggested [12] that there may be a gossamer condensate in all materials that have metal → bad metal phase transitions similar to the $\alpha \rightarrow \gamma$ transition in Ce. The presumed reason that these materials are metallic rather than bulk superconductors is that the quasi-particle excitation gap is very small, not just at certain points, but over sizable areas of the Fermi surface. In this paper we argue that the "gossamer metal" picture can actually explain both the quantum critical behavior of Np/Pu/Am as well as the volume collapse phase transitions in rare earth materials. In particular, we will argue that in both cases the observed phase transitions are consequences of rapid yet continuous changes in the gossamer condensate wave function. In the case of the rare earth volume collapse transitions the continuity of the wave function is suggested by the existence of a critical point at ~ 600 $^0$K similar to an ordinary liquid-gas critical point. Evidently the $\alpha$ and $\gamma$ phases of Ce are connected in much the same way a gas can be continuously transformed into a liquid by going around the critical point.

In the case of the various phases of Pu and its neighbors Np and Am the materials melt before reaching a classical critical point. However, the phase transitions are typically martensitic, i.e. they can be visualized as resulting from displacements in a unit cell. Although martensitic transformations are generically first order, they can be accompanied by continuous changes in electronic structure [13]. In particular, Anderson and Blount [13] point out that as a result of the coupling of the lattice and electronic degrees of freedom a martensitic transformation can lead to the appearance of local electric fields. This might explain, for example, the appearance of the complex lattice structures in the actinides. It is interesting in this connection that in the case of Sm the volume discontinuity vanishes, and there is a proliferation of complex phases on the high pressure side of the transition. Actually the possibility that the phase transitions in Pu may be related to the volume collapse transitions in the rare earths has been discussed for a long time [14].

The density of conduction electrons in a BCS superconductor is equal to twice the density of paired carriers. In a gossamer metal, on the other hand, the condensate density is not related to the number of conduction electrons in a simple way [12]. In general pairing correlations produce a peak in the density of occupied states near to the Fermi surface. Correlation effects which reduce the effect of pairing will therefore decrease the density of quasi-particles with energies near to the Fermi surface. Thus a decrease in the condensate density at fixed pressure would be expected to be accompanied by a decrease in the number of conduction electrons. This could result in a dramatic decrease in the conductivity. The volume collapse transitions have often been interpreted as Mott metal-insulator transitions [14,15]. However, the existence of a critical point for the $\alpha \rightarrow \gamma$ transition in Ce implies that these transitions cannot be metal-insulator transitions, which are always first order phase transitions. In the gossamer condensate picture the large decrease in electrical conductivity associated with the volume collapse is due to a decrease in the number of carriers with energies near to the Fermi surface [12], while the effective mass of the carriers remains nearly constant. The density discontinuity for the volume collapse transition will vanish at the critical point, which in some cases, e.g. Sm, can be near to room temperature.

As it happens a simple band theory of the $\alpha \rightarrow \gamma$ transition in Ce leads to van der Waals-like behavior for the volume vs. pressure curves of the elemental rare earths [16]. When pairing correlations are important, the usual Hartree-Fock theory should be replaced by the Bogolubov-Hartree-Fock theory. In this case the volume vs. pressure curves of the elemental rare earths may simply mirror the fact that a quantum fluid at zero temperature can exhibit Van der Waals-like behavior as a function of some parameter in the Hamiltonian. Of course pressure vs. volume curves of real materials are typically measured at room temperature, and so one might question whether the room temperature volume collapse phenomena can be identified with the behavior of a quantum fluid at zero temperature.

Fortunately CeThLa alloys provide us with examples of volume collapse transitions where the transition can be made to occur at very low temperatures by appropriate choice of La doping [17] or with the application of a large magnetic field [18]. In the case of $Ce_{0.9-x}Th_{0.1}La_x$ the volume discontinuity appears to vanish or at least become very weak as the critical temperature approaches zero. Furthermore, it has very recently been observed that near $x \sim 0.14$ the temperature dependence of the specific heat and resistance is similar to that observed near to quantum critical points of $f$-electron materials [19]. A distinguishing feature though is that for a La doping on either side of $x_c$ the material appears to be a normal heavy fermion Fermi liquid. Near the critical point the magnetic susceptibility appears to have a broad peak as the La doping $x$ is varied. This is exactly the kind of behavior that is expected for a gossamer metal near to a carrier concentration where anti-ferromagnetism fluctuations become important due to the "magnetic resonance mode" [20]. It is worth noting that this behavior is not consistent with the sudden appearance of Kondo screening on one side of the transition, which has been a popular interpretation for the volume collapse in the rare earths [21]. An alternative to the Kondo collapse model would naturally be attractive in the case of Pu where there are for sure no local moments.

What remains to be discussed is the physical origin of the gossamer pairing in the actinides. In the presence of lattice distortions that lead to internal electric fields, spin orbit effects can lead to the spontaneous appearance of spin currents. If the length scale associated with the appearance of these spin currents is smaller than the screening length then the charge carriers become quasi-localized [22]. Soliton-like localization of charge carriers does not by itself imply that charge carriers are paired. However, it turns out that spin orbit localized carriers look like magnetic monopoles, and so it is very natural for carriers with opposite spin to be paired (Fig. 2). Actually simple symmetry considerations imply that an electric field can lead to spin pairing only if spin orbit interactions are important. Pairing of itinerant *f*-electrons due to spin orbit effects is a plausible candidate for explaining both the "gossamer metal" properties of the rare earth and actinide materials, as well as the recently discovered examples of exceptional superconductivity in actinide materials.

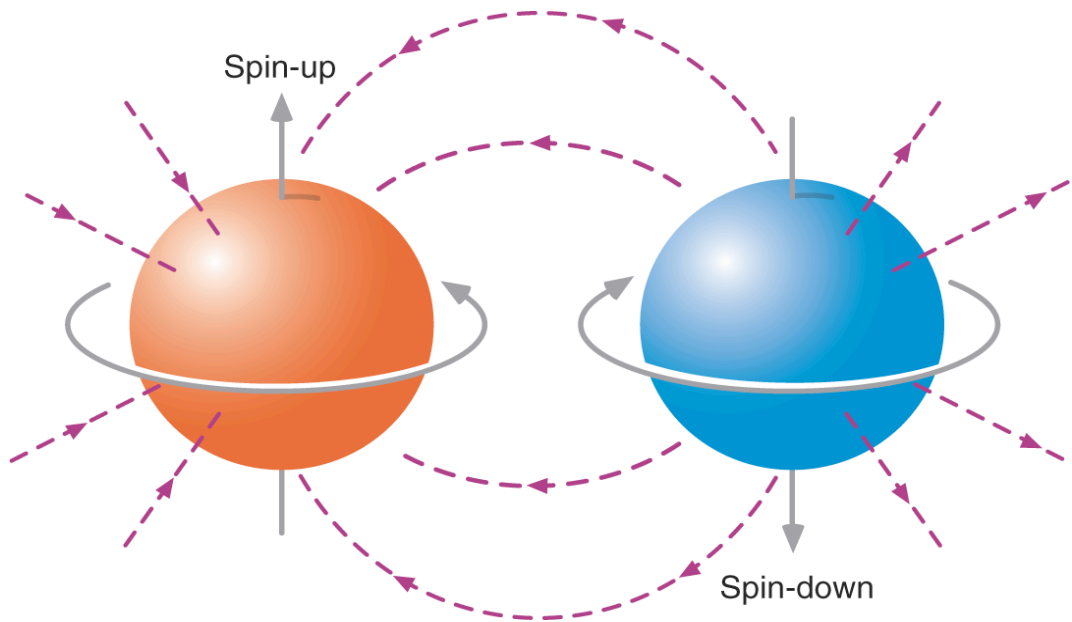

Fig.2 Pairing of spin orbit localized carriers

The authors would like to thank Claude Guet, Jason Lashley, Gil Lonzarich, Brian Rainford, Peter Riseborough, David Santiago, Jim Smith, and Joe Thompson for helpful conversations and suggestions. This work was performed in part under the auspices of the U.S. Department of Energy by University of California Lawrence Livermore National Laboratory under contract No. W-7405-Eng-48.